\documentclass[a4paper,10pt]{article}
\usepackage{graphicx}
\usepackage{enumitem}
\usepackage{subfigure}
\usepackage{amssymb}
\usepackage{amsmath}
\usepackage{bbm}
\usepackage{color}
\usepackage{cite}

\begin{document}

\title{Noncommutative Geometry and the Primordial Dipolar Imaginary Power Spectrum}

\author{Pankaj Jain and 
Pranati K. Rath} 
\maketitle

\begin{center}
{Dept. of Physics, Indian Institue of Technology Kanpur, Kanpur - 208016, India}

\medskip
Email:pkjain@iitk.ac.in,  pranati@iitk.ac.in
\end{center}

\begin{abstract}
We argue that an anisotropic dipolar imaginary primordial power spectrum is
possible within the framework of noncommutative space-times.
We show that such a spectrum provides a good description of the observed
dipole modulation in CMBR data. We extract the corresponding power spectrum from data. The dipole modulation is related to
 the observed hemispherical anisotropy in CMBR data, which
might represent the first signature of quantum gravity.
\end{abstract}

\maketitle

\section{Introduction}

The cosmic microwave background radiation
(CMBR) shows a hemispherical power anisotropy 
\cite{Eriksen2004,Eriksen2007,Hansen2009,Hoftuft2009,Erickcek2008,Planck2013a,Paci2013,Schmidt2013,Akrami2014}, which  
can be parametrized as,
\begin{equation}
 {\bigtriangleup T}(\hat n) = g(\hat n) \left(1+A \hat \lambda \cdot \hat n \right)
\label{eq:dipole_mod}
\end{equation}
where $ g(\hat n)$ is an isotropic and Gaussian random field,
  $\hat \lambda$ the preferred direction and $A$ the amplitude of
anisotropy. This model implies a 
 dipole modulation \cite{Gordon2005,Gordon2007,Prunet2005,Bennett2011} 
of the CMBR temperature field. 
The WMAP five year data leads to $A=0.072\pm0.022$ with the 
dipole direction, 
$(\theta,\phi) =(224^o,112^o)\pm24^o$ for $l \le 64$
in the galactic coordinates 
\cite{Eriksen2004,Eriksen2007,Hansen2009,Erickcek2008,Paci2013,Schmidt2013}. 
This anisotropy has been confirmed by
PLANCK \cite{Planck2013a} with amplitude and direction similar to those 
found in WMAP. 
The hemispherical anisotropy has also been probed at 
multipoles higher than $64$ \cite{Hansen2009,Hoftuft2009}.
The signal is found to be absent at $l\sim 500$
 \cite{Donoghue2005,Hanson2009} and also not seen in the 
 large scale structures 
\cite{Hirata2009,Fernandez2013}. These observations may be accommodated
in a model which proposes a 
scale dependent power spectrum \cite{Erickcek2009}, such that the effect
is negligible at high$-l$.

Many theoretical models,
such as, 
\cite{Berera2004,ACW2007,Boehmer2008,Jaffe2006,Koivisto2006,Land2006,
Bridges2007,Campanelli2007,Ghosh2007,Pontez2007,Koivisto2008,Kahniashvili2008,
Carroll2010,Watanabe2010,Chang2013a,Anupam2013a,Anupam2013b,Cai2013,Liu2013,
Chang2013b,Chang2013c,Ghosh2014},
have been proposed which 
aim to explain the observed hemispherical anisotropy as well as other
signals of anisotropy seen in data 
\cite{Hutsemekers1998,Jain1999,Costa2004,Ralston2004,Schwarz2004,Singal2011,Tiwari2013}. 
An interesting possibility is that there might have been a
phase of anisotropic expansion at very early time. The inflationary Big Bang
cosmology is perfectly consistent with such an evolution.
The anisotropic modes, generated during this
early phase may later re-enter the horizon
\cite{Aluri2012,Pranati2013a} and lead to the observed signals. 

In this paper our objective is to determine a primordial power spectrum
which may lead to dipole modulation and hence hemispherical anisotropy.
It is possible to find a power spectrum based on an inhomogeneous model
\cite{Erickcek2008,Hirata2009,Carroll2010,Gao2011,Mcdonald2014,Aslanyan:2014mqa} 
which is consistent \cite{Rath2014} with the observed temperature anisotropy.
However it is not clear how an anisotropic model might lead to a dipole
modulation. The simplest model that one might construct
leads to quadrupolar and not dipolar modulation \cite{Rath2014}. 
The basic problem can
be understood by considering the two point correlations in real space.
Let $\tilde\delta(\vec x)$ be the density fluctuations in real space.
Their two point correlation function, $F(\vec\Delta,\vec X)$, 
may be expressed as,
\begin{equation}
F(\vec\Delta,\vec X) = \langle\tilde\delta(\vec x) \tilde\delta(\vec x')\rangle 
\end{equation} 
where, $\vec \Delta= \vec x-\vec x'$ and
$\vec X= (\vec x+\vec x')/2$. 
We are interested in a correlation which is anisotropic and hence depends on
$\vec \Delta$ besides the magnitude $\Delta \equiv|\vec \Delta|$. 
It is clear from the definition of the correlation
function that, in a classical framework, it
must satisfy,
\begin{equation}
\langle\tilde\delta(\vec x) \tilde\delta(\vec x')\rangle 
=
\langle\tilde\delta(\vec x') \tilde\delta(\vec x)\rangle 
\label{eq:symmetry}
\end{equation} 
 Hence it can only be an even function of $\vec \Delta$. The simplest
anisotropic function is, therefore,
\begin{equation}
F(\vec\Delta,\vec X) = f_1(\Delta) + B_{ij}\Delta_i\Delta_j f_2(\Delta)
\end{equation} 
where $B_{i,j}$, $i,j=1,2,3$ are parameters. It is clear that such a model
cannot give rise to a dipole modulation, which requires a term linear
in $\Delta_i$.   

In this paper we argue that in a noncommutative space-time, a term linear
in $\Delta_i$ is permissible. The power spectrum that we are interested
in is applicable at very early time, perhaps even the time when quantum
gravity effects were not negligible. At that time, we cannot assume that 
space-time is commutative \cite{Doplicher1994,Connes1994,Madore1999,Landi1997,Bondia2001}. 
Its noncommutativity may be expressed as,
\begin{equation}
[\hat x_\mu,\hat x_\nu] = i\theta_{\mu\nu}
\label{eq:basiccommutator}
\end{equation}
where, $\theta_{\mu\nu}$ are parameters 
and the coordinate functions, $\hat x_\mu(x)$, 
depend on the choice of coordinate system. 
In a particular coordinate system, we may set, 
\begin{equation}
\hat x_\mu(x) = x_\mu 
\end{equation}
In \cite{Akofor2008} the authors assume that this prefered system
is the comoving coordinate system. In general, the noncommutativity
can appear quite complicated in different systems. 
The effect of noncommutativity on cosmology has been considered 
earlier \cite{Greene2001,Lizzi2002,Brandenberger2002,Huang2003,Brandenberger2003,Bal2008,Fatollahi2006,Fatollahi2006a,Akofor2008,Akofor2009,Shiraishi2014,Cai2014}.
However its relationship with dipole modulation has not been pointed out 
so far.

In \cite{Rath2014}, we have determined the power spectrum
corresponding to an inhomogeneous model
and shown that its spectral index is consistent with zero.
In the present paper we determine the power spectrum of an anisotropic model
based on noncommutative space-time.

\section{Correlations induced by Dipole Modulation}
In this section we review the correlations between different multipoles
which are induced by the dipole modulation model, Eq. \ref{eq:dipole_mod}. 
We may expand the CMBR temperature as,  
\begin{equation}
{\bigtriangleup T}(\hat n) = \sum_{lm}a_{lm}Y_{lm}(\hat n)
\end{equation}
where, $a_{lm}$ are the spherical harmonic coefficients. 
Their two point correlation function  
may be expressed as,
\cite{Pranati2013b}, 
\begin{equation}
\langle{a_{lm}a^*_{l'm'}}\rangle = \langle{a_{lm}a^*_{l'm'}}\rangle_{iso} +\langle{a_{lm}a^*_{l'm'}}\rangle_{dm} 
\label{eq:corrdm}
\end{equation}
where, 
\begin{eqnarray}
 \langle{a_{lm}a^*_{l'm'}}\rangle_{iso} &=& 
                                            C_l\delta_{ll'}\delta_{mm'}\nonumber
\\
 \langle{a_{lm}a^*_{l'm'}}\rangle_{dm} &=& A\left(C_{l'}+
                                            C_l\right)\xi^{0}_{lm;l'm'}
\label{eq:corr_aniso}
\end{eqnarray}
Here $C_l$ is the standard angular power spectrum, 
$\langle{a_{lm}a^*_{l'm'}}\rangle_{dm}$ is the contribution due to the
anisotropic dipole modulation and
\begin{eqnarray}
  \xi^{0}_{lm;l'm'}=\delta_{m',m}\Bigg[\sqrt{\frac{(l-m+1)(l+m+1)}{{(2l+1)}{(2l+3)}}}\delta_{l',l+1}\nonumber\\
+\sqrt{\frac{(l-m)(l+m)}{{(2l+1)}{(2l-1)}}}\delta_{l',l-1}\Bigg]\ .
\label{eq:xillprime}
\end{eqnarray}
Hence the model leads to correlations between multipoles, $l$ and $l+1$.
We define the statistic \cite{Pranati2013b}, 
\begin{equation}
S_H(L) = \sum_{l = l_{min}}^{L} C
 \frac{l(l+1)}{2l+1}\sum_{m = -l}^{l} a_{lm}a^*_{l+1,m}
\label{eq:SH}
\end{equation}
We maximize the statistic by varying over the direction parameters.
The resulting statistic is labelled as $S_{H}^{data}$.  
This provides a measure of the signature of anisotropy seen in data.
This can be compared with a theoretical power spectrum model in order to 
fix its parameters.

\section{Anisotropic Power Spectrum}
The relationship between the temperature fluctuations, 
${\bigtriangleup T(\hat n)}$, and the 
 primordial density perturbations, $\delta(\vec k)$, can be expressed as, 
\begin{equation}
\frac{\bigtriangleup T}{T_0}(\hat n) = \int d^{3}k \sum_{l}(-i)^{l}(2l+1)\delta(k)\Theta_l(k) P_l(\hat k \cdot \hat n)
\end{equation} 
where $P_l(\hat k \cdot \hat n )$ are the 
Legendre polynomials, 
\begin{equation}
P_l(\hat n \cdot \hat n') = \frac{4\pi}{2l+1}\sum_{m} Y_{lm}(\hat n)Y^*_{lm}(\hat n')\,, 
\label{eq:Legendre}
\end{equation}
and $\Theta_l(k)$ the transfer function. Here we  
assume an
approximate form of the transfer function,
 $\Theta_l(k) =\frac{3}{10}j_l(k\eta_0)$
\cite{Gorbunov}, where $j_l$ is the spherical Bessel function. 

We next propose the following form of the 
anisotropic power spectrum in real
space,
\begin{equation}
F(\vec \Delta) = f_1(\Delta) + \hat\lambda\cdot\vec\Delta f_2(\Delta)
\label{eq:FDelta}
\end{equation}
where $\hat\lambda$ represents the preferred direction
and $f_1$ and $f_2$ depend only on the magnitude $\Delta$. 
Such a form is generally not permissible since the correlation function 
must satisfy Eq. \ref{eq:symmetry}.
However this does not follow in noncommutative space-time \cite{Akofor2008}. 
In this case the relevant quantity is the deformed quantum field. Let 
$\phi_0(\vec x,t)$ be a self-adjoint scalar field. The deformed quantum
field is defined as \cite{Akofor2008},
\begin{equation}
\label{eq:twistedfield}
\varphi_{\theta} = \varphi_{0}\; \textrm{e}^{\frac{1}{2}\overleftarrow{\partial} \wedge P}
\end{equation} 
where
\begin{equation}
\overleftarrow{\partial} \wedge P \equiv \overleftarrow{\partial}_{\mu}\theta^{\mu \nu}P_{\nu}.
\end{equation} 
For a deformed field,
\begin{equation}
\Phi_{\theta}({\bf x}, t) \Phi_{\theta}({\bf x}', t') \neq
\Phi_{\theta}({\bf x}', t') \Phi_{\theta}({\bf x}, t)
\end{equation}  
for space like separations \cite{Akofor2008}. 
Hence Eq. \ref{eq:symmetry} does not follow and a correlation function,
Eq. \ref{eq:FDelta}, which
depends linearly on $\vec \Delta$, is permissible. 

The correlation function  
of the Fourier transform, 
$\delta(\vec k)$, of $\tilde\delta(\vec x)$ may be expressed as, 
\begin{equation}
\langle \delta(\vec k)\delta^*(\vec k')\rangle = \int {d^3X\over (2\pi)^3}
{d^3\Delta \over (2\pi)^3} 
e^{i(\vec k+\vec k')\cdot \vec \Delta/2}
e^{i(\vec k-\vec k')\cdot \vec X} F(\vec\Delta,\vec X)
\label{eq:corr1}
\end{equation} 
Using the model given in Eq. \ref{eq:FDelta}, we obtain,
\begin{equation}
\langle \delta(\vec k)\delta^*(\vec k')\rangle = \delta^3(\vec k-\vec k')
\int d^3\Delta 
e^{i(\vec k+\vec k')\cdot \vec \Delta/2}
F(\vec\Delta)
\label{eq:corr2}
\end{equation} 
This leads to,
\begin{equation}
\langle{\delta(\vec k)\delta^*(\vec k')}\rangle = \delta^3(\vec k-\vec k')P(k)[1+i(\hat k \cdot \hat \lambda )g(k)]
\label{eq:corr3}
\end{equation} 
where the delta function arises due to spatial translational invariance and
 $P(k)$ is the standard power spectrum,
\begin{equation}
P(k) = k^{n-4}A_{\phi}/(4\pi)
\label{eq:poweriso}
\end{equation} 
Here we set the parameters,  
 $n =1$ and $A_{\phi} = 1.16\times 10^{-9}$ \cite{Gorbunov}. 
In Eq. \ref{eq:corr3}, $g(k)$ is a real function which 
 depends only on the magnitude $k=|\vec k|$ and 
represents the violation of statistical isotropy. 
A more detailed fit is postponed to future work.

We may compare this for the power spectrum obtained in 
\cite{Akofor2009}, for the commutator,
\begin{equation}
{1\over 2}[\phi_\theta(\vec x, \eta), \phi_\theta(\vec y, \eta)]_-
\equiv {1\over 2} \left(\phi_\theta(\vec x, \eta) \phi_\theta(\vec y, \eta)
- \phi_\theta(\vec y, \eta)\phi_\theta(\vec x, \eta)   \right)
\label{eq:commutator}
\end{equation}
where $\eta$ is the conformal time. 
In Fourier space the correlator is given by Eq.  17 of
\cite{Akofor2009}, reproduced here for convenience,
\begin{eqnarray}
{1\over 2}<0|[\phi_\theta(\vec k, \eta), \phi_\theta(\vec k', \eta)]_-|0>\Bigg|_{\rm
horizon\ crossing}
\nonumber\\ = (2\pi)^3P(k)i\sinh(H\vec \theta^0\cdot\vec k)\delta(\vec k+\vec k') 
\label{eq:commutator1}
\end{eqnarray}
where we correct a crucial typographical error in \cite{Akofor2009} regarding
the presence of the imaginary $i$. 
In this equation $P(k)$ is the standard power spectrum, given in Eq. 
\ref{eq:poweriso} and $\vec \theta^0=(\theta^{01}, \theta^{02}, \theta^{03})$
are three parameters. 
The argument of the Dirac delta function is
$(\vec k+\vec k')$ instead of 
$(\vec k-\vec k')$  
in Eq. \ref{eq:corr2} 
since here we take the correlation between 
$\phi_\theta(\vec k, \eta)$ and $\phi_\theta(\vec k', \eta)$
instead of 
$\phi_\theta(\vec k, \eta)$ and $\phi^\dagger_\theta(\vec k', \eta)$.
In the limit of small anisotropy parameters, $\vec \theta^0$, we can expand 
the sinh function and keep only the leading order term. Comparing with
Eq. \ref{eq:corr3}, we identify,
\begin{equation}
g(k) = Hk|\vec \theta^0|
\label{eq:gknoncomm}
\end{equation}
and the direction, $\hat\lambda=\hat \theta^0$. 
The precise form of the correlation predicted within the framework
of noncommutative geometry is model dependent.
In particular, the basic equation, Eq. \ref{eq:basiccommutator}, depends 
on the choice of 
coordinates which obey this simple relationship. 
Here we don't confine
ourselves to a particular model and instead extract the anisotropic
power directly from data. 
For this purpose, we assume the following parametrization of $g(k)$,
\begin{equation}
g(k) = g_0 (k\eta_0)^{-\alpha}\ .
\label{eq:modelgk}
\end{equation}
 where $g_0$ and $\alpha$ are parameters.

We next compute the two point temperature correlations, 
\begin{eqnarray}
 \langle{{\bigtriangleup T}(\hat n) {\bigtriangleup T(\hat n')}}\rangle = T_0^2
\int d^{3}k\sum_{l,l'=0}^{\infty}(-i)^{l-l'}\nonumber\\
\times (2l+1)(2l'+1)\Theta_l(k) \Theta_{l'}(k)\nonumber\\
\times P_l(\hat k \cdot \hat n)P_{l'}(\hat k \cdot \hat n')P_{iso}(k)[1+ig(\hat k \cdot \hat \lambda )]
\end{eqnarray}
Setting z-axis as
the preferred direction, we obtain $\hat k \cdot \hat \lambda = \cos{\theta} $.
The correlations of the spherical harmonic coefficients can be expressed as,
\begin{equation}
 \langle{a_{lm}a^*_{l'm'}}\rangle=\int d\Omega_{\hat n} d\Omega_{\hat n'} Y^*_{lm}(\hat n)Y_{l'm'}(\hat n')\langle 
{\bigtriangleup T(\hat n)}{\bigtriangleup T(\hat n')}\rangle
\end{equation}
We finally obtain,
\begin{equation}
 \langle{a_{lm}a^*_{l'm'}}\rangle = \langle{a_{lm}a^*_{l'm'}}\rangle_{iso}+\langle{a_{lm}a^*_{l'm'}}\rangle_{aniso}\,,
\end{equation}
where,
\begin{equation}
\langle{a_{lm}a^*_{l'm'}}\rangle_{iso} = (4\pi)^2 \frac{{9T_0}^2}{100}\delta_{ll'}\delta{mm'}\int_0^{\infty} k^2dk j_l^2(k\eta_0)P_{iso}(k)\,, 
\end{equation}
\begin{equation}
\langle{a_{lm}a^*_{l'm'}}\rangle_{aniso} =(-i)^{l-l'+1} (4\pi)^2 
\frac{9{T_0}^2}{100}G_{ll'}
  \xi^{0}_{lm;l'm'}\,, 
\label{eq:theorycorr}
\end{equation}
$\xi^{0}_{lm;l'm'}$ is defined in Eq. \ref{eq:xillprime} and
\begin{equation}
G_{ll'} = \int_0^{\infty} k^2dk P(k)j_{l}(k\eta_o)j_{l'}(k\eta_0)
g(k)\, .
\end{equation}
Using Eq. \ref{eq:modelgk}, we obtain,
\begin{equation}
G_{ll'} = {g_0A_\phi\over 4\pi} \int_0^{\infty} {ds\over s^{1+\alpha}} j_{l}(s)j_{l'}(s)
\ .
\end{equation}
Hence 
the anisotropic power spectrum, Eq. \ref{eq:FDelta}, leads to a 
correlation between $l$ and $l\pm 1$. 
This allows us to obtain the theoretical prediction of the statistic, 
$S_H(L)$, which can be compared to to $S_{H}^{data}$ in order to determine
the best fit value of power spectrum parameters, $g_0$ and $\alpha$.

\section{Data Analysis} 
We use the cleaned CMB map, ILC, based on WMAP 9 year data \cite{WMAP9yrILC} 
(hereafter WILC9)
and
SMICA, provided by the PLANCK team \cite{Planck2013b}. 
We use the KQ85 and 
CMB-union mask in order to eliminate foreground contaminated regions
for the WMAP and PLANCK data respectively.
We generate a full sky map from the masked data by filling the
 masked portion with simulated data. The simulated CMB maps contain
contribution due to the dipole modulation. We first generate a
full sky CMB map by using 
isotropic and Gaussian random field. This map is generated at  
high resolution with $Nside = 2048$. The resulting map is multiplied with
the dipole modulation term, $(1+A\hat\lambda\cdot \hat n)$ in order to
generate a full sky map which has same properties as the real data.
The data from this map is used to fill the gaps in the real map. 
This data map
is downgraded to a lower resolution with $Nside = 32$ after 
applying appropriate 
Gaussian beam to smooth the mask boundary \cite{Pranati2013b}.
Hence any breaks that might be introduced at the
boundary of the masked region get eliminated. 
We also use the SMICA in-painted map, in which the in-painting
procedure \cite{Abrial2008,Inoue2008}
has been used to reconstruct the masked regions,
 provided by the PLANCK team.

In order to determine the power spectrum parameters,
 we first set $\alpha=0$ and determine the 
best fit value of $g_0$ over the entire 
multipole range $2\le l\le 64$. 
The maximum value of the
statistic, $S_H(L)$, in this multipole 
range is determined by maximizing over the
preferred direction parameters. The resulting value of the statistic
depends on the random realization used to fill the masked regions.  
Hence the  
 maximum value of $S_H(L)$ and $(\theta,\phi)$,
are obtained by taking an average over 100 full sky data maps.  
Here $(\theta,\phi)$ are the direction parameters in polar coordinates.
The resulting statistic is compared with theoretical prediction
in order to determine the best fit value of $g_0$ with the 
constraint, $\alpha=0$.

We next determine the best fit values
of both $g_0$ and $\alpha$ by spliting data into three multipole bins,
 $l=2-22,23-43,44-64$. In this case we find it convenient to fix
the direction parameters to be same as those obtained over the
entire multipole range. As shown in \cite{Rath2014}, these show some 
dependence on the multipole bin, but the dependence
 is relatively mild and we ignore it
for present analysis.

\section{Results}

For the entire multipole range,
$2\le l\le 64$ the best fit value of $g_0$ is found to be,
$g_0=0.32\pm 0.08$ and $g_0=0.30\pm 0.08$ for WILC9 and SMICA 
respectively.  
Here we have assumed that the spectral index $\alpha=0$. Hence the 
function, $g(k)$ is equal to a constant, $g_0$. 
We have verified that the results obtained for the case of the SMICA
in-painted map 
 are in good agreement with those for SMICA and WILC9.

\begin{figure}[!th]
\centering
\includegraphics[scale=0.60,angle=0]{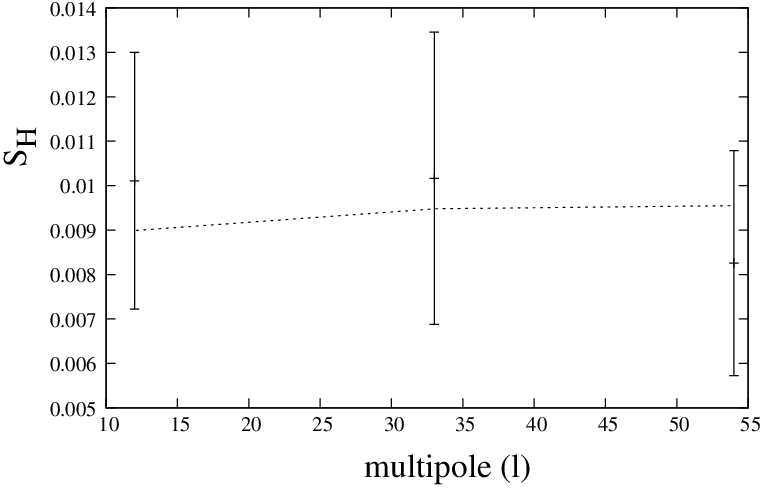}
\caption{The statistic, $S_{H}^{data}$, as a function of the multipole $l$
for WILC9. 
Here the statistic in the three bins is extracted by fixing the direction
parameters to be equal to the mean direction over the entire multipole range.
The dotted line corresponds to the theoretical 
fit corresponding to $\alpha=0, g_0=0.32\pm 0.06$. 
}
\label{fig:SHfit}
\end{figure}

We next extract the function, $g(k)$, using data in the three
multipole bins, $l=2-22,23-43$ and $l=44-64$. We parametrize it 
 in terms of
$g_0$ and the spectral index $\alpha$.  
Setting $\alpha=0$, the best fit value of $g_0$ is found to be,
$g_0=0.32\pm 0.06$ with $\chi^2=0.45$ for WILC9
and $g_0=0.30\pm 0.05$ with $\chi^2=0.41$ for SMICA. 
Hence we find that 
 a zero spectral index for the anisotropic 
part of the power spectrum provides a good fit to data. 
The resulting fit is shown 
in Fig. \ref{fig:SHfit} as the dotted line.
Allowing a non-zero value of $\alpha$ we find that the $1\sigma$ limit
on this parameter is, $-0.13<\alpha<0.15$ and $-0.16<\alpha<0.19$ for
WILC9 and SMICA respectively.

\section{Conclusion}
We show that an anisotropic power spectrum model, derived on the basis of
noncommutative geometry provides a description of the observed hemispherical
anisotropy. This anisotropy can be parametrized in terms of a dipole modulation
model, which leads to correlations among the multipoles corresponding
to $l$ and $l+1$. The noncommutative anisotropic power spectrum model also
leads  
to such a correlation. The anisotropic power spectrum is parameterized
by the function, $g(k)$. We first fit the data by assuming that $g(k)$ is
a constant equal to $g_0$. 
We determine the value of $g_0$ by making first making a fit over
the entire multipole range, $2-64$. The best fit value is found to 
be $g_0=0.32\pm0.08$.  
We next assume a power law form of $g(k)= g_0(k\eta_0)^{-\alpha}$ and extract the corresponding
amplitude, $g_0$ and spectral index $\alpha$ by making a fit over the
three multipole bins, $l=2-22,23-43$ and $l=44-64$. 
Setting $\alpha=0$, the best fit leads to $\alpha = 0.32\pm 0.06$ for WILC9. 
This leads to a good fit to data with $\chi^2=0.45$. Hence the
data suggests that the anisotropic power, $g(k)$, is independent of $k$.
Furthermore we find the one sigma limit on $\alpha$ to be,
$-0.13<\alpha<0.15$ for WILC9.  

We conclude that 
the observed hemispherical anisotropy might represent the first observational
signature of noncommutative geometry and hence of quantum gravity.

{\bf Acknowledgments:}
We acknowledge the use of Planck data available from NASA’s LAMBDA site
(http://lambda.gsfc.nasa.gov). Some of the results in this paper have been derived using
the Healpix package \cite{Gorski2005}. We are extremely grateful to A. P. Balachandran for a very useful discussion on noncommutative space-times. We also
thank James Zibin for a useful communication.

\end{document}